\documentclass[aps,prc,twocolumn,groupedaddress,showpacs]{revtex4}

\usepackage{graphicx}

\begin{document}

\title{ Heavy-residue isoscaling as a probe of the symmetry energy of hot fragments }
\author{G. A. Souliotis}
\author{D. V. Shetty}
\author{A. Keksis}
\author{E. Bell}
\author{M. Jandel}
\author{M. Veselsky}
\thanks{Institute of Physics of the Slovak Academy of Sciences, Bratislava, Slovakia.}
\author{S. J. Yennello}

\affiliation{Cyclotron Institute, Texas A\&M University, College Station, TX 77843}

\date{\today}

\nopagebreak

\begin{abstract}

The isoscaling properties of isotopically resolved projectile
residues from peripheral collisions  of $^{86}$Kr (25 MeV/nucleon),
$^{64}$Ni (25 MeV/nucleon) and $^{136}$Xe (20 MeV/nucleon) beams  on
various target pairs are employed to probe the symmetry energy
coefficient of the nuclear binding energy.
 The  present study focuses on heavy projectile fragments produced in peripheral and
semiperipheral collisions near the onset of multifragment emission
(E$^{*}$/A = 2--3 MeV). For these fragments, the measured average
velocities are  used to extract excitation energies. The excitation
energies, in turn, are used to estimate the temperatures of the
fragmenting quasiprojectiles in the framework the Fermi gas model.
The isoscaling analysis of the fragment yields provided the
isoscaling parameters $\alpha$ which, in combination with
temperatures and isospin asymmetries provided  the symmetry energy
coefficient of the nuclear binding energy of the hot fragmenting
quasiprojectiles. The extracted values of the symmetry energy
coefficient at this excitation energy range (2--3 MeV/nucleon) are
lower than the typical liquid-drop model value $\sim $ 25 MeV
corresponding to ground-state nuclei and show a monotonic decrease
with increasing excitation energy. This result is of importance in
the formation of hot nuclei in heavy-ion reactions and in hot
stellar environments such as supernova.


\end{abstract}



\maketitle

\section{Introduction}

The study of the nuclear symmetry energy  is currently a topic of
intense theoretical and experimental work. It is well established
that the symmetry energy plays a central role in a variety of
astrophysical phenomena including the structure and evolution of
neutron stars and the dynamics of supernova explosions
\cite{ASTE05,LEE96,BET90,PET95,LAT00,HIX03,JStone1,ZBotvina0}. In
addition, the symmetry energy determines the nuclear structure of
neutron-rich or neutron deficient  rare isotopes
\cite{OYA98,BRO00,HOR01,FUR02}.

 The symmetry energy at normal nuclear density has been
obtained from a number of many-body approaches \cite{
KHO96,ZUO99,BRA85,PEA00,DIE03} and from nuclear mass systematics
\cite{MYE66,POM03,JStone2,Zmass-review}. However, its values at
densities below or above the normal nuclear density are not
adequately constrained \cite{Shetty05,Shetty04,BAO05}. Indeed, the
experimental determination of the symmetry energy and its density
dependence is a challenging scientific endeavor. Information on the
symmetry energy can be gleaned from the determination of the neutron
skins of neutron-rich nuclei \cite{OYA98}, from elastic scattering
on neutron-rich nuclei \cite{KHO05} and from heavy-ion collisions.
For the latter, a great deal of effort is currently devoted  in
order to identify observables sensitive to the nuclear symmetry
energy and its density dependence (see, e.g.,
\cite{BAOreview,BAObook,BAO05,Ono03,ImQMD}).

One important observable in heavy ion collisions is the fragment
isotopic composition investigated with the recently developed
isoscaling approach \cite{Tsang1}. The isoscaling  approach attempts
to isolate the effects of the nuclear symmetry energy in the
fragment yields, thus allowing a direct study of the role of
this term of the nuclear binding energy in the formation of hot
fragments. 
Isoscaling refers to a general exponential relation between the
yields of a given fragment from two reactions that differ only in
their isospin asymmetry (N/Z) \cite{Tsang1,Tsang2,Botvina}.
In particular, if two reactions, 1 and 2, lead to primary fragments
having approximately the same temperature but different isospin
asymmetry, the ratio R$_{21}$(N,Z) of the yields of a given fragment
(N,Z) from these primary fragments exhibits an exponential
dependence on the neutron number N and the atomic number Z of the
form:
\begin{equation}
   R_{21}(N,Z) = C \exp(\alpha N + \beta Z)
\end{equation}
where $\alpha$ and $\beta$ are the scaling parameters and C is a
normalization constant. This scaling behavior has been observed in a
very broad range of reactions including evaporation
\cite{Tsang1,Botvina}, fission \cite{MVffiso,Friedman},
deep-inelastic reactions \cite{Tsang1,GSiso,GSnz} and
multifragmentation
\cite{Tsang1,Botvina,Shetty03,MViso,Geraci,Trautmann,ZKHS}. In the
initial studies of isoscaling, it was shown that the  isoscaling
parameters are almost unaffected by the sequential decay of the
primary fragments, due to possibly similar de-excitation paths of
the two primary fragments, thus they could provide information on
the early stage of fragment formation.
In particular, within the statistical framework, the isoscaling
parameter $\alpha$ is linearly
related to the symmetry energy  coefficient 
of the fragment binding energy \cite{Tsang1,Botvina,Ono03,Ono04}.
This relation provides the key connection of the measured isoscaling
parameter with the symmetry energy coefficient.

In the present work, the isoscaling properties of isotopically
resolved projectile residues from peripheral collisions  of
$^{86}$Kr (25 MeV/nucleon), $^{64}$Ni (25 MeV/nucleon) and
$^{136}$Xe (20 MeV/nucleon) beams  on a variety of target pairs are
employed to probe the symmetry energy coefficient of the nuclear
binding energy. The collection and complete characterization of the
residues in terms of their atomic number Z, mass number A and
velocity  has been performed with  the use of two magnetic
separators: the MARS recoil separator and the Superconducting
Solenoid Line at Texas A\&M University.  In this work, apart from
isotopically resolved yields, the velocities of the fragments are
obtained with high resolution and are used to provide information on
the excitation energy (and temperature) of the primary fragments.
The isoscaling  parameters $\alpha$ along with temperatures and
isospin asymmetries yielded the values of the nuclear symmetry
energy in the  excitation energy range 2--3 MeV/nucleon.
The paper is organized as follows. In Section II, a brief
description of the experimental devices, the measurements and the
data analysis is given. In Section III, the isotopic scaling of the
fragment yields and the velocity distributions are presented. In
Section IV, the systematics of the isoscaling parameter $\alpha$
with respect to isospin asymmetry is presented and used
to get the symmetry energy. Finally, conclusions from the present
study are summarized in Section V.

\section{Experimental Methods and Data Analysis}

The present studies were  performed at the Cyclotron Institute of
Texas A\&M University using two different devices: the MARS recoil
separator and the Superconducting Solenoid Line. Below we give a
concise description of the measurements with each of these devices.
\subsection{Measurements with the MARS recoil separator}
The reactions of $^{86}$Kr with  $^{64,58}$Ni and $^{124,112}$Sn
were studied with the MARS recoil separator. The general isoscaling
analysis of these data has been reported recently \cite{GSiso}.
Nevertheless, for a complete presentation of the heavy-residue
isoscaling approach to probe the symmetry energy, we briefly
summarize the method below. A 25 \hbox{MeV/nucleon} $^{86}$Kr beam
from the K500 superconducting cyclotron, with a current of $\sim$1
pnA, interacted with isotopically enriched targets of $^{64}$Ni,
$^{58}$Ni and $^{124}$Sn, $^{112}$Sn. The reaction products entered
the MARS spectrometer \cite{MARS} having  an angular acceptance of 9
msr and momentum acceptance of 4\%. The primary beam struck the
target at 4.0$^{o}$ relative to the optical axis of the
spectrometer.
Fragments were accepted in the polar angular range
2.7$^{o}$--5.4$^{o}$. This angular range lies inside the grazing
angle $\theta_{gr}$=6.5$^{o}$ of the Kr+Sn  reactions and mostly
outside the grazing angle $\theta_{gr}$=3.5$^{o}$ of the Kr+Ni
reactions   at 25 MeV/nucleon \cite{Wilcke}. (The spectrometer angle
setting was chosen to optimize the production of very neutron-rich
fragments from the Kr+Sn systems whose detailed study  has been
reported in \cite{GSprl}.)  An Al stripper foil (1 mg/cm$^2$) was
used  to reset to equilibrium the ionic charge states of the
projectile fragments. MARS optics \cite{MARS} provides one
intermediate dispersive  image and one (final) achromatic image
(focal plane). At the focal plane, the fragments were collected in a
5$\times$5 cm  two-element ($\Delta $E, E)  Si detector telescope.
Time of flight was measured between two PPACs (Parallel Plate
Avalanche Counters) positioned at the dispersive image and at the
focal plane, respectively, and separated by a distance of 13.2  m.
The PPAC at the dispersive image was also  X--Y  position sensitive
and used  to record  the position of the fragments. The horizontal
position, along with NMR measurements of the field of the MARS first
dipole, provided the magnetic rigidity, $B\rho$,  of the particles.

The determination of the atomic number Z was based on the energy
loss of the particles in the $\Delta E$ detector 
 and their velocity. The ionic charge $q$ of the particles entering MARS
was obtained from the total energy E$_{tot}$=$\Delta$E + E, the
velocity and the magnetic rigidity. The measurements of Z and q had
resolutions of 0.5 and 0.4 units (FWHM), respectively. Since the
ionic charge is an integer, we assigned integer values of q for each
event by putting windows ($\Delta q=0.4$) on each peak of the q
spectrum. Using the magnetic rigidity and velocity measurement, the
mass-to-charge A/q ratio  of each ion was obtained with a resolution
of 0.3\%. Combining the q determination with the A/q measurement,
the mass A was obtained as: $A = q_{int} \times A/q $ \, (q$_{int}$
is the integer ionic charge) with a resolution  (FWHM) of 0.6 A
units. Combination and  normalization of the data at the various
magnetic rigidity settings of the spectrometer (in the range
1.3--2.0 T\,m), summation over all ionic  charge states
and, finally,  normalization for beam current  and target thickness,
provided fragment yield distributions with respect to Z, A  and
velocity. Further details of the analysis procedure can be found in
\cite{GSplb,GSnim}.
The yield distributions, summed over velocities,  were used to
obtain the fragment yield ratios $ R_{21}(N,Z) =
Y_{2}(N,Z)/Y_{1}(N,Z) $ employed  in the present isoscaling studies.

\subsection{Measurements with the Superconducting Solenoid Line}
The heavy-residue work with the Superconducting Solenoid Line shares
many  similarities with the previously described work with MARS and
it is described below. The complete (two-stage) Superconducting
Solenoid Line (Fig. 1) consists of the 7-Tesla superconducting
solenoid (``BigSol'') of the University of Michigan (first stage)
\cite{Tom1,Tom2} and a large-bore quadrupole triplet (second stage).
The whole separator line is also referred to as the ``BigSol'' line.
Details of the development of the line and plans/progress towards
producing rare isotopes in deep-inelastic collisions are provided in
\cite{GS-bigsol,GSnim}.


    \begin{figure}[t]                                        

    \includegraphics[width=0.45\textwidth, height=0.15\textheight ]{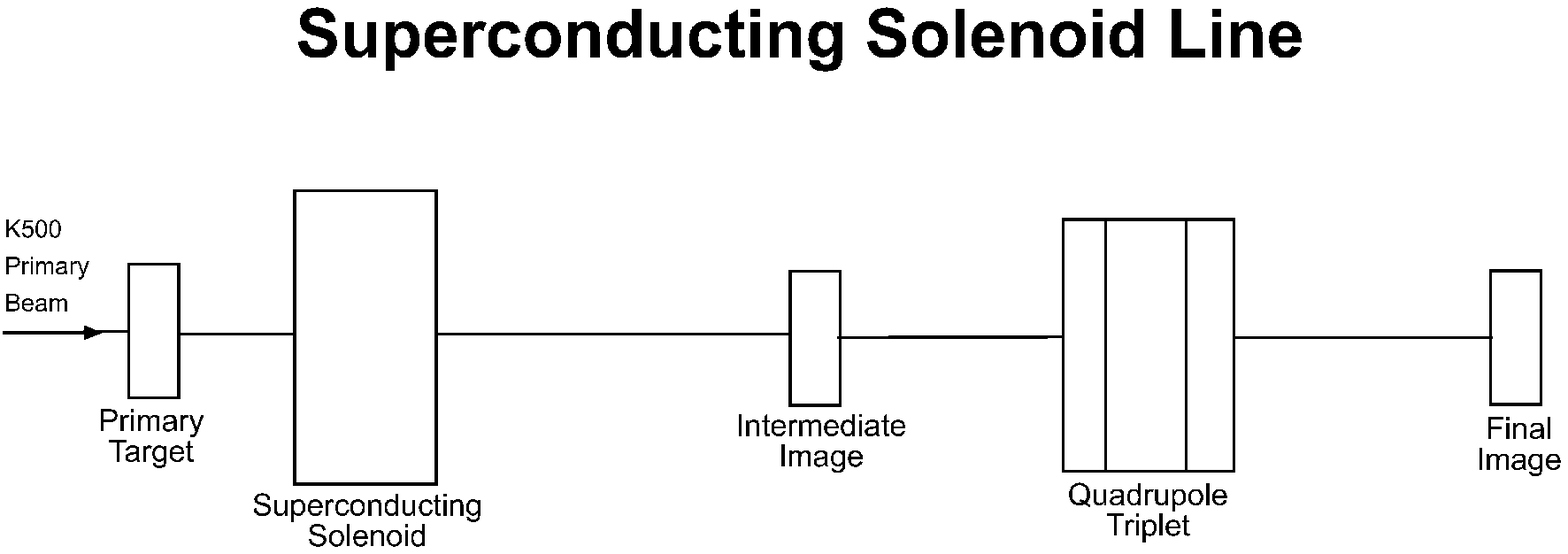}

    \caption{ Schematic diagram of the Superconducting Solenoid Line (BigSol) used in
              the present heavy residue isoscaling studies (see text)}
    \label{vel}
    \end{figure}


In the present isoscaling studies with BigSol, first, the reactions
of a $^{64}$Ni (25 MeV/nucleon) beam with targets of $^{64,58}$Ni,
$^{124,112}$Sn and $^{208}$Pb, $^{232}$Th were studied. Second, the
reactions of a $^{136}$Xe (20 MeV/nucleon) beam with targets of
$^{64,58}$Ni, $^{124,112}$Sn and $^{197}$Au, $^{232}$Th were
measured.  The typical beam current was $\sim$1 pnA. The primary
beam, after hitting the target, was collected on an on-axis blocker
located $\sim$30 cm after the target. The beam blocker along with a
circular aperture at this location defined the angular acceptance of
the line to be 1.5$^{o}$--3.0$^{o}$. The fragments then passed
through a stripper foil, traversed the solenoid and were focused at
the intermediate image (Fig. 1). At this location, a  magnetic
rigidity (or momentum-over-charge p/q) acceptance of $\sim$4\% was
defined with another circular aperture. Subsequently,  the fragments
were transported through a 7.5 meter line and focused with the aid
of the quadrupole triplet at the end of the line (final image).
Time-of-flight was provided between two X-Y position sensitive
PPACs, one at the intermediate image and the other at the final
image. A silicon detector array similar to the one used in the MARS
measurements provided $\Delta$E, E information, which, combined with
time-of-flight (with resolution $\sim$0.5\%) provided Z and A
determination (with resolutions of 0.5 and 0.6 units, respectively,
for Ni-like fragments).

It should be noted that, in contrast to the MARS measurements, here
the mass determination was based solely on total energy and
time-of-flight. The reason is that BigSol does not provide
high-resolution p/q dispersion.  The Solenoid Line is not a magnetic
spectrometer in the classical sense \cite{spectrograph}. Its
elements are only focusing elements. There is radial p/q dispersion
at the intermediate image due to the solenoid (and, of course, at
the final image, due to the quadrupole triplet) as a consequence of
the variation of the location of the focus  with p/q. This
dispersion is also a function of the initial angle \cite{Tom1}. The
measurement of the radial distance of the fragments at the
intermediate image (emerging, as stated previously, in the initial
angular range of 1.5$^{o}$--3.0$^{o}$), combined with the value of
the central magnetic field of the solenoid, provided a p/q
determination with a resolution of $\sim$2\%. Even though this
resolution is not useful to improve the mass determination (as was
done in the MARS data \cite{GSplb}), it was adequate to specify the
charge state q of the ions.

As in the case of the MARS data, a series of runs at overlapping
magnetic rigidity settings of the line in the range 1.1--1.6 T\,m
for the $^{64}$Ni (25MeV/nucleon) data and 1.0--1.5 T\,m  for the
$^{136}$Xe (20 MeV/nucleon) data were performed. The data were
normalized and appropriately combined, following the procedure
described in \cite{GSplb}. Summation over all ionic charge states
and, finally, normalization for beam current and target thickness,
provided the fragment yield distributions with respect to Z, A  and
velocity. As in the case of the MARS data, the fragment yield
distributions were summed over velocities and used to obtain the
yield ratios for isoscaling.

\section{Isoscaling and excitation energy data from the various reactions}

From the measured  fragment yield distributions, we construct the
ratio $R_{21}(N,Z) = Y_{2}(N,Z) / Y_{1}(N,Z)$  of yields of a given
projectile fragment (N,Z) from reactions 2 and 1 using the
convention that index 2 refers to the more neutron-rich system and
index 1 to the less neutron-rich one. The results of the reactions
with each of the various beams are given in the following
subsections.
\subsection{ $^{86}$Kr (25MeV/nucleon) data (MARS) }
Fig. 2 shows the yield ratios  R$_{21}$(N,Z) as a function of
fragment neutron number N for selected isotopes for the Kr+Ni
reactions (top panel) and the Kr+Sn reactions (bottom panel). The
different isotopes are shown by alternating filled and open symbols
for clarity. For each element, an exponential function of the form
C$exp$($\alpha$ N) was fitted to the data and is shown in Fig. 2 for
the selected elements. In the semi-log representation, the straight
lines for each element are parallel up to Z$\sim$34 for Kr+Ni and up
to Z$\sim$28 for Kr+Sn. For heavier fragments from the Kr+Sn system,
the fits to the data show gradual decrease in the slopes with
increasing Z, a behavior that has been shown to manifest incomplete
N/Z equilibration for the most peripheral events for this reaction
whose projectile-like fragments were observed inside the grazing
angle \cite{GSiso,GSnz}.

    \begin{figure}[h]                                        
    \includegraphics[width=0.47\textwidth, height=0.50\textheight ]{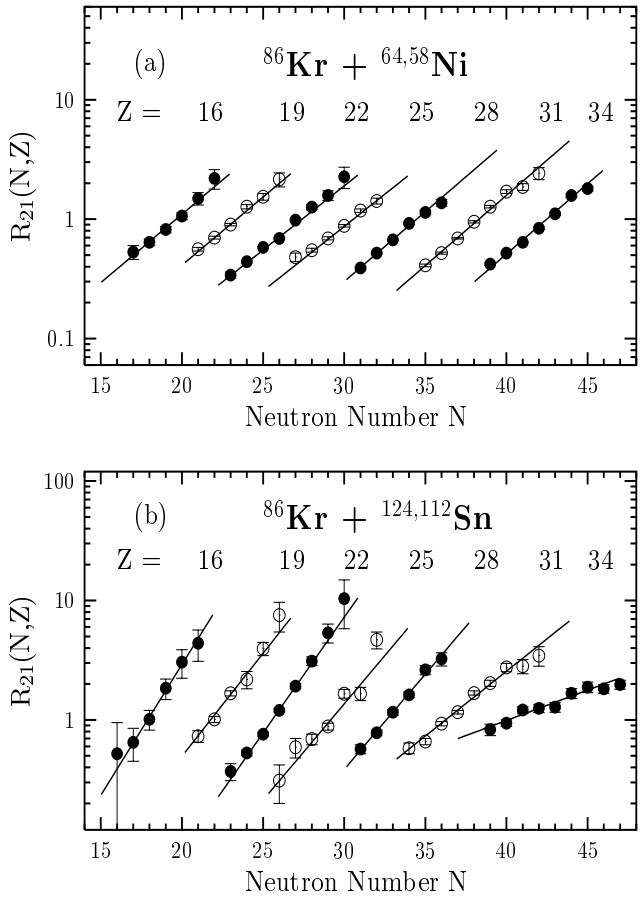}

    \caption{ (MARS data)
              Yield ratios $ R_{21}(N,Z) = Y_{2}(N,Z)/Y_{1}(N,Z) $ of  projectile residues
              from  the reactions of $^{86}$Kr (25 MeV/nucleon) with $^{64,58}$Ni  (a), and
              $^{124,112}$Sn (b) with respect to N for the Z's indicated.
              The data are given by alternating filled and open circles, whereas the lines are
              exponential fits.
            }
    \label{isoscale_mars}
    \end{figure}
    \begin{figure}[h]                                        
    \includegraphics[width=0.47\textwidth, height=0.40\textheight ]{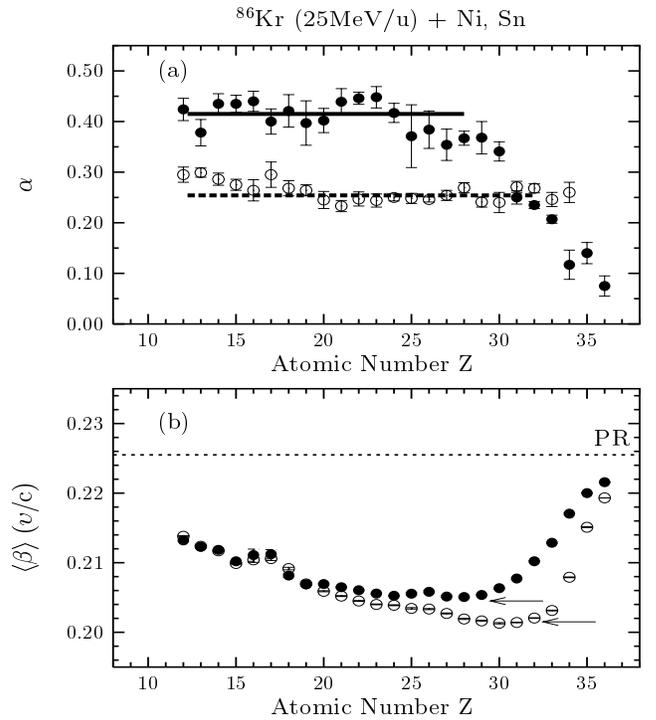}

    \caption{(MARS data)
             (a) Isoscaling parameter $\alpha$ as a function of Z for
                 projectile residues from  the reactions of  $^{86}$Kr (25 MeV/nucleon) with
                 $^{64,58}$Ni (open circles) and $^{124,112}$Sn (closed
                 circles). The straight lines are constant value fits for each system.
             (b) Average velocity versus atomic number Z  correlations  for projectile residues
                 from  the reactions of $^{86}$Kr (25 MeV/nucleon) with
                 $^{64}$Ni (open symbols) and $^{124}$Sn (closed
                 symbols).
                 The dashed line (marked ``PR'') gives the velocity of the projectile, whereas
                 the arrows indicate the minimum average residue velocities observed.
             }
    \label{alpha_vel_mars}
    \end{figure}
In  Fig. 3a,  we present  the slope parameters $\alpha$ of the
exponential  fits (obtained as described for Fig. 2) as a function
of Z.
For the Kr+Ni reactions (open symbols), the slope parameter $\alpha$
is constant in the whole range Z=12--36 at an average  value
$\alpha$=0.254.
For the Kr+Sn reactions, the parameter $\alpha$ is roughly constant
with  an average  value of 0.415 in the range Z=12--28 and then it
decreases  for Z$>$28.
In the case of the Kr+Ni systems under the present experimental
conditions (observation outside the grazing angle), isoscaling holds
essentially in the whole range of observed fragments.
To obtain information about  the excitation energy of the primary
fragments from the present reactions, we will employ the correlation
of the  measured velocity with the atomic number \cite{GSiso}.
Fig. 3b  presents the average velocities of the fragments as a
function of  Z. Open symbols correspond to the $^{86}$Kr+$^{64}$Ni
reaction and closed symbols to the $^{86}$Kr+$^{124}$Sn reaction. In
this figure, we observe that for fragments  close to the projectile,
the velocities are slightly below that of the projectile,
corresponding to very peripheral, low-excitation energy events. A
monotonic decrease of velocity with decreasing Z is observed,
indicative of lower impact parameters, and thus, higher excitation
energies.
For the  $^{86}$Kr+$^{124}$Sn reaction (closed symbols in Fig. 3b),
the descending velocity--Z  correlation continues  down to
Z$\sim$28; for lower Z's, the velocity starts increasing  with
decreasing Z. A minimum velocity for Z$\sim$28 can be understood by
assuming that these residues originate from primary fragments with a
maximum observed excitation energy.
Fragments with Z near the projectile down to Z$\sim$28 originate
from evaporative type of deexcitation  which does not modify, on
average, the emission direction of the residues. Thus, the residue
velocities provide information on the excitation energy.
Residues with lower Z  arise from primary fragments undergoing
cluster emission and/or multifragmentation and the velocity of the
inclusively measured fragments is not monotonically related to the
excitation energy.
For the  $^{86}$Kr+$^{64}$Ni reaction, a similar behavior is
observed.  However, the decreasing velocity--Z correlation is
observed  down to  Z$\sim$32. We remind that fragments from this
reaction were measured mostly outside the grazing angle, so that
they correspond to more damped  collisions, in such a way that the
final residues receive a larger recoil during the deexcitation and
appear  within this angular range. For Z$\sim$30--32, we observe a
minimum velocity  and  for lower Z's an increase of the velocity
with decreasing Z analogous to the $^{86}$Kr+$^{124}$Sn reaction.
We mention that the average velocities from the reactions with the
neutron-poor targets are, within the experimental uncertainties,
almost the same as the corresponding from the reactions with the
neutron-rich targets for both pairs of systems and are not shown in
Fig. 3b. For both reactions, the ascending  part of the velocity vs
Z correlation for the lower part of the Z range is primarily due to
the combined effect of angle and  magnetic rigidity selection
imposed by the spectrometer.

Employing  the observed minimum  velocities  for the   Kr+Ni and
Kr+Sn reactions and, furthermore, assuming two-body  kinematics, we
can estimate the total excitation energy of the quasiprojectile --
quasitarget system employing standard mass tables \cite{Moller}. In
regards to the sharing of excitation energy, a reasonable assumption
for peripheral/semiperipheral collisions is equal division
\cite{Madani,TokeEx,MV_SiSn} at relatively low kinetic energy
losses, E$_{loss}$. This assumption was employed in our previous
analysis of these isoscaling data \cite{GSiso}. In the present work,
we use  a more appropriate prescription in agreement with  the
 experimentally observed transition of the excitation energy sharing
 from the equal division limit (at low E$_{loss}$) to the thermal limit
that may be attained  near the maximum of the kinetic energy loss,
E$_{loss,max}$.
Specifically, we estimate the fraction of the excitation energy of
the quasi-projectile assuming a linear evolution, with respect to
E$_{loss}$/E$_{loss,max}$, from the equal division limit to the
thermal equilibrium limit.
Following the above procedure, we can estimate an average excitation
energy per nucleon for the hot quasiprojectile fragments of E$^*$/A
= 2.4 MeV for the Kr+Ni system and E$^*$/A = 2.0 MeV for the Kr+Sn
system (Table I). We note that, for these systems, the present
estimates are close to the value E$^*$/A = 2.2 MeV obtained under
the assumption of equal excitation energy division employed in our
previous work \cite{GSiso}.


\subsection{ $^{64}$Ni (25MeV/nucleon) data (BigSol) }

In Figs. 4 and 5, we show  the data for the $^{64}$Ni (25
MeV/nucleon) reactions obtained with  the BigSol line. The
presentation of the data follows a line similar to that of the MARS
data (Figs. 2 and 3). In Figs. 4a,b, a general isoscaling behavior
is seen  in both the Ni+Ni (grazing angle $\theta_{gr}$=3.8$^{o}$)
and Ni+Sn ($\theta_{gr}$=6.5$^{o}$) reactions essentially in the
whole range of elements measured in the magnetic rigidity range
1.1--1.6 T\,m. A slight decrease of the isoscaling parameter
$\alpha$ is discernible for near-projectile elements (Fig. 5a)
possibly due to incomplete N/Z equilibration. Constant value fits to
the $\alpha$ parameter vs Z in the range Z=12--24 yielded
$\alpha$=0.250 and 0.324 for Ni+Ni and Ni+Sn respectively (Table I).


    \begin{figure}[h]                                        
    \includegraphics[width=0.47\textwidth, height=0.65\textheight ]{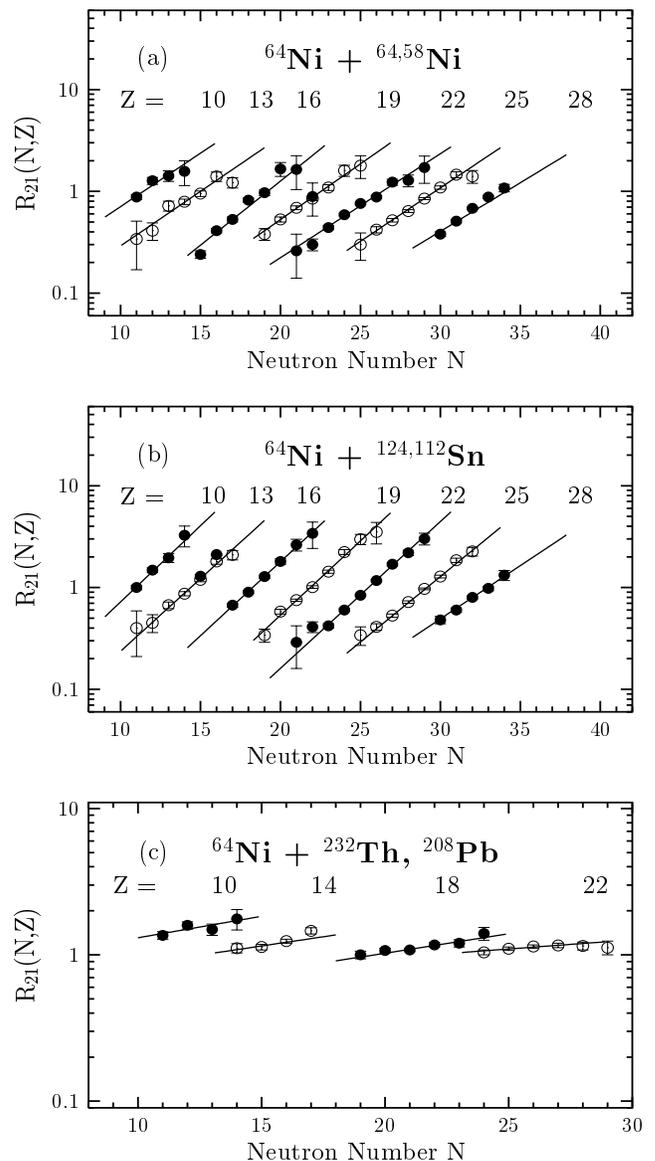}

    \caption{ (BigSol data)
              Yield ratios $ R_{21}(N,Z) = Y_{2}(N,Z)/Y_{1}(N,Z) $ of  projectile residues
              from  the reactions of  $^{64}$Ni (25 MeV/nucleon) with
              $^{64,58}$Ni (a),
              $^{124,112}$Sn (b), and
              $^{232}$Th,$^{208}$Pb (c)
              with respect to N for the Z's indicated.
              The data are given by alternating filled and open circles, whereas the lines are
              exponential fits.
            }
    \label{isoscale_bigsol_1}
    \end{figure}


    \begin{figure}[h]                                        
    \includegraphics[width=0.47\textwidth, height=0.40\textheight ]{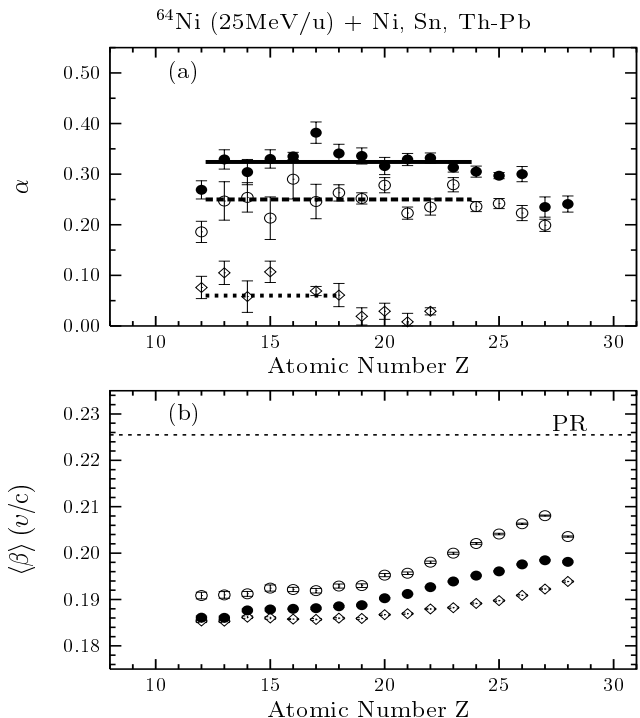}

    \caption{(BigSol data)
             (a) Isoscaling parameter $\alpha$ as a function of Z for
                 projectile residues from  the reactions of $^{64}$Ni (25 MeV/nucleon)
                 with
                 $^{64,58}$Ni (open circles),
                 $^{124,112}$Sn (closed circles), and
                 $^{232}$Th,$^{208}$Pb (open diamonds).
                 The straight lines are constant value fits for each system.
             (b) Average velocity versus atomic number Z  correlations  for projectile residues
                 from  the reactions of $^{64}$Ni(25 MeV/nucleon) with
                 $^{64}$Ni  (open symbols),
                 $^{124}$Sn (closed symbols), and
                 $^{232}$Th (open diamonds).
                 The dashed line (marked ``PR'') gives the velocity of the projectile.
             }
    \label{alpha_vel_bigsol_1}
    \end{figure}
For the $^{64}$Ni + $^{232}$Th, $^{208}$Pb pair of reactions
($\theta_{gr}$=9.5$^{o}$), we also observe isoscaling despite the
small difference of $\sim$2.5\% in the N/Z of the two systems
(compared to, e.g., the $\sim$10\% N/Z difference in the reaction
pair of $^{64}$Ni + $^{124}$Sn,$^{112}$Sn). This limiting case shows
the sensitivity of the isoscaling signal to the N/Z of the reacting
systems. A constant value fit to the $\alpha$ parameter data of the
Ni+Th,Pb pair in the region Z=14--18 (Fig. 5a) gives, as expected, a
small value $\alpha$=0.060 (Table I).

The average velocities of the three $^{64}$Ni reactions are shown in
Fig. 5b (only the neutron-rich systems are presented). As we see in
the figure, near-projectile residues with velocities close to that
of the projectile were not collected in these measurements, mainly
due to magnetic rigidity selection (and, in addition,  due to angle
selection $\Delta\theta$=1.5--3.0$^{o}$, particularly for the
Ni+Th,Pb systems with $\theta_{gr}$=9.5$^{o}$).  In a manner similar
to the MARS data, we employed the observed minimum velocities to
extract average excitation energies for the fragmenting Ni-like
quasiprojectiles resulting in a common value of E$^{*}$/A = 2.9 MeV
for the three pairs of the $^{64}$Ni reactions (Table I).


\subsection{ $^{136}$Xe (20 MeV/nucleon) data (BigSol) }

Finally, in Figs. 6 and 7,  we present the data for the reactions of
the $^{136}$Xe (20 MeV/nucleon) beam on the three pairs of targets
$^{64,58}$Ni, $^{124,112}$Sn and  $^{232}$Th, $^{197}$Au obtained
with the BigSol line. Again, the presentation of the data is similar
to that of the preceding sections. Fig. 6 shows the isoscaling
behavior of the yields and the fits to selected elements. Fig. 7a
shows the variation of the isoscaling parameter $\alpha$  as a
function of Z. The $\alpha$ values for Xe+Ni and Xe+Sn are displaced
vertically for viewing. Despite the relatively large fluctuation in
the data points, the data were fitted with straight lines in the
region Z=14--46 yielding the values: 0.129, 0.193,  and 0.096 for
the three reaction pairs, respectively (Table I).


    \begin{figure}[h]                                        
    \includegraphics[width=0.47\textwidth, height=0.65\textheight ]{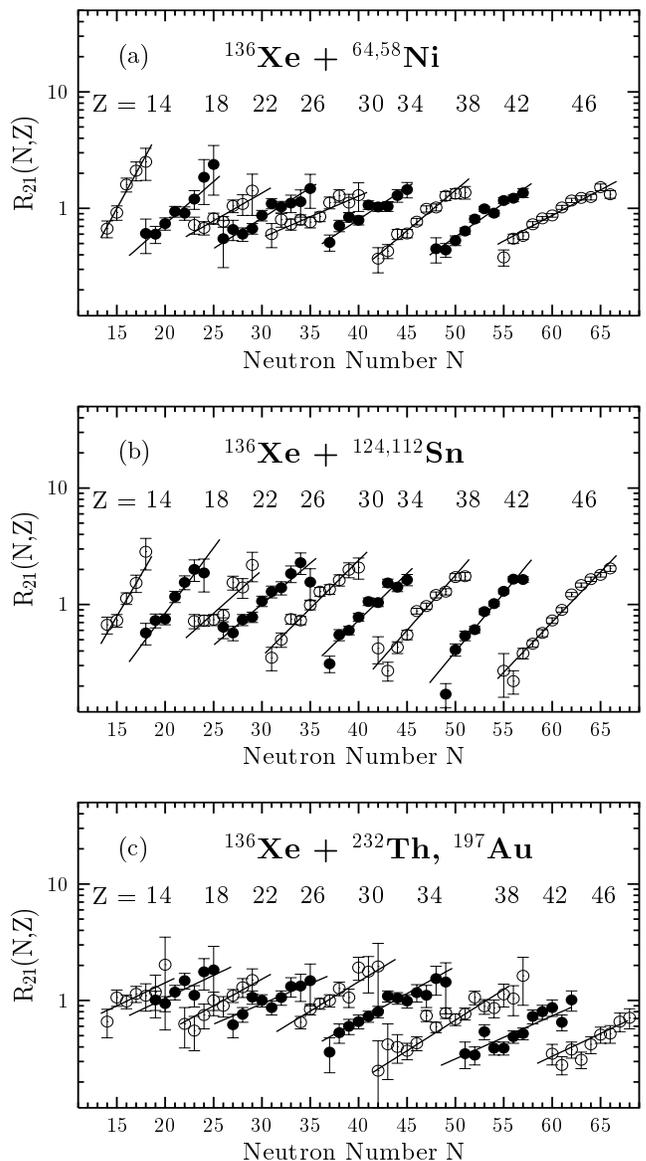}

    \caption{ (BigSol data)
              Yield ratios $ R_{21}(N,Z) = Y_{2}(N,Z)/Y_{1}(N,Z) $ of  projectile residues
              from  the reactions of  $^{136}$Xe (20 MeV/nucleon) with
              $^{64,58}$Ni (a),
              $^{124,112}$Sn (b), and
              $^{232}$Th,$^{197}$Au (c)
              with respect to N for the Z's indicated.
              The data are given by alternating filled and open circles, whereas the lines are
              exponential fits.
            }
    \label{isoscale_bigsol_2}
    \end{figure}


    \begin{figure}[h]                                        
    \includegraphics[width=0.47\textwidth, height=0.40\textheight ]{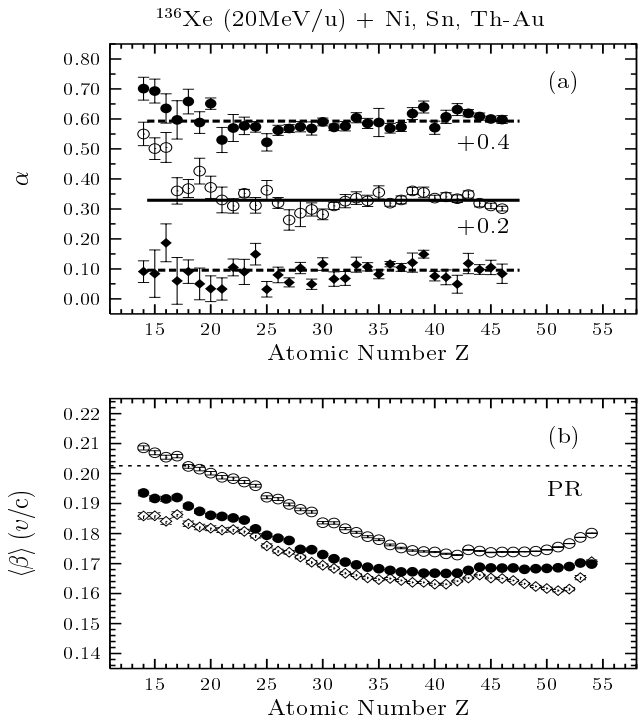}

    \caption{(BigSol data)
             (a) Isoscaling parameter $\alpha$ as a function of Z for
                 projectile residues from  the reactions of $^{136}$Xe (20 MeV/nucleon)
                 with
                 $^{64,58}$Ni (open circles),
                 $^{124,112}$Sn (closed circles), and
                 $^{232}$Th,$^{208}$Pb (closed diamonds).
                 The values of the first two systems are displaced  vertically by +0.4 and +0.2
                 units, respectively, for viewing. The straight lines are constant value fits for each system.
             (b) Average velocity versus atomic number Z  correlations  for projectile residues
                 from  the reactions of $^{136}$Xe (20 MeV/nucleon) with
                 $^{64}$Ni  (open symbols),
                 $^{124}$Sn (closed symbols), and
                 $^{232}$Th (open diamonds).
                 The dashed line (marked ``PR'') gives the velocity of the projectile.
             }
    \label{alpha_vel_mars}
    \end{figure}

Fig. 7b shows the average velocities of the three $^{136}$Xe
reactions (again, only the neutron-rich systems are presented). The
grazing angles for these systems are 3.9$^{o}$, 6.3$^{o}$ and
10.0$^{o}$ respectively. The  angle and magnetic rigidity selection
were such that near-projectile residues with velocities close to
that of the projectile were not collected, as in the case of the
$^{64}$Ni reactions. However, the kinematical conditions were such
that lighter fragments (below Z=35) show an ascending velocity
behavior with decreasing Z analogous to the one observed in the
Kr+Ni,Sn reactions (Fig. 3b). Using the observed minimum velocities,
average excitation energies for the fragmenting Xe-like
quasiprojectiles were extracted  resulting again in a common value
of E$^{*}$/A = 2.5 MeV for the three pairs of the $^{136}$Xe
reactions.

\section{ Interpretation  of results and discussion }
Before proceeding to the analysis and interpretation of the present
results on heavy fragment isoscaling, we present a summary of the
reactions and the relevant parameters in Table I.
The difference in the isotopic  composition, expressed as the proton
 fraction squared, $(Z/A)^{2}$, for each pair of systems is given along with the
excitation energies (obtained from residue velocities), as well as
the temperatures (calculated with the Fermi gas model and the
expanding mononucleus model \cite{Sobotka}, as discussed below). The
isoscaling parameters $\alpha$ and, finally,  the extracted values
of the symmetry energy coefficient are summarized.
%
%
%
\begin{table}[t]                     
\caption{ Summary of reaction pairs studied in this work (see text),
          along with parameters relevant to the isoscaling analysis:
          $\Delta (Z/A)^{2}$ difference in (Z/A)$^{2}$ of the two reactions of each pair.
          E$^{*}$/A (MeV): excitation energy per nucleon of the corresponding primary
          quasiprojectiles. K (MeV):  inverse level density
          parameter.
          T$_{F}$, T$_{m}$(MeV): temperature estimates obtained from the Fermi gas model and the
          expanding  mononucleus model \cite{Sobotka}.
          $\alpha$: isoscaling parameter.
          $ c \equiv \alpha / \Delta (Z/A)^{2}$ : reduced isoscaling
          parameter.
          C$_{sym,F}$, C$_{sym,m}$ (MeV): symmetry energy coefficient of the nuclear
          binding energy obtained using the two temperature
          determinations  T$_{F}$ and  T$_{m}$
          respectively, as mentioned above.
          Errors
          (one standard deviation) of the measured quantities
          are given in parentheses.
          }
\vspace{0.5cm}
\rm

\footnotesize

\begin{center}
\begin{tabular}{cccccccccc} \hline \hline
Reaction &$\Delta (Z/A)^{2}$& E$^{*}$/A & K  & T$_{F}$& T$_{m}$ &
 $\alpha$ & c & C$_{sym,F}$ & C$_{sym,m}$ \\ \hline
         &                  &           &      &     &        &        &      &        &          \\
Kr+Sn    &  0.0209          & 2.0       & 11.5 & 4.8 &  3.9   &  0.415 & 19.9 &  23.9  &  19.4    \\
         &                  &(0.1)      &      &     &        & (0.010)& (0.5)&  (0.6) &  (0.5)   \\
         &                  &           &      &     &        &        &      &        &          \\
Kr+Ni    &  0.0154          & 2.4       & 11.8 & 5.3 &  4.2   &  0.254 & 16.5 &  21.9  &  17.2    \\
         &                  &(0.1)      &      &     &        & (0.005)& (0.3)&  (0.5) &  (0.4)   \\ \hline
         &                  &           &      &     &        &        &      &        &           \\
Ni+Ni    &  0.0193          & 2.9       & 11.7 & 5.8 &  4.3   & 0.250  & 13.0 &  18.9  &  14.0     \\
         &                  &(0.1)      &      &     &        &(0.007) & (0.4)&  (0.6) &  (0.4)    \\
         &                  &           &      &     &        &        &      &        &           \\
Ni+Sn    &  0.0240          & 2.9       & 11.7 & 5.8 &  4.3   & 0.324  & 13.5 &  19.6  &  14.5     \\
         &                  &(0.1)      &      &     &        &(0.005) & (0.3)&  (0.5) &  (0.4)    \\
         &                  &           &      &     &        &        &      &        &           \\
Ni+Th--Pb&  0.0046          & 2.9       & 11.7 & 5.8 &  4.3   & 0.060  & 13.0 &  18.9  &  14.0     \\
         &                  &(0.1)      &      &     &        &(0.010) & (2.2)&  (3.2) &  (2.2)    \\ \hline
         &                  &           &      &     &        &       &      &        &           \\
Xe+Ni    &  0.0106          & 2.5       & 13.5 & 5.8 &  4.5   & 0.129 & 12.2 &  17.7  & 13.7      \\
         &                  &(0.1)      &      &     &        &(0.006)& (0.6)&  (0.9) & (0.7)     \\
         &                  &           &      &     &        &       &      &        &           \\
Xe+Sn    &  0.0159          & 2.5       & 13.5 & 5.8 &  4.5   & 0.193 & 12.1 &  17.6  & 13.6      \\
         &                  &(0.1)      &      &     &        &(0.005)& (0.3)&  (0.5) & (0.4)     \\
         &                  &           &      &     &        &       &      &        &           \\
Xe+Th--Au&  0.0064          & 2.5       & 13.5 & 5.8 &  4.5   & 0.096 & 15.0 &  21.8  & 16.9      \\
         &                  &(0.1)      &      &     &        &(0.005)& (0.8)&  (1.2) & (0.9)     \\
         &                  &           &      &     &        &       &      &        &           \\
\hline \hline
\end{tabular}
\end{center}
\end{table}
\subsection{ Determination of temperature }
In order to estimate  the temperature of the fragmenting
quasiprojectiles using the measured average  excitation energies, we
first employ the simple Fermi gas relationship of the form  $ E^* =
\frac{A}{K} T^2 $, with T the tempretature and K the inverse level
density parameter.
It is well known that for the non-interacting Fermi gas model the
value of the inverse level density parameter is K $\simeq$16 (e.g.
\cite{Shlomo1}),  whereas the experimental data are consistent with
lower values of K dependent on both the excitation energy and the
mass range \cite{JBN1}. For the present systems in the 2--3
MeV/nucleon excitation range, values in the vicinity of K=12--13 are
consistent with the experimental systematics \cite{JBN1}. Given the
expected mass and excitation energy dependence of K, we decided to
use values of K following the model of Shlomo and Natowitz
\cite{Shlomo2,Shlomo3}. This model determines the nuclear level
density within the framework of the Fermi gas incorporating the
effects of the finite size of the nucleus, the contributions of the
continuum states and the temperature and density dependence of the
nucleon effective mass. For the hot quasiprojectiles (Kr-like,
Ni-like and Xe-like) of the present work, the values of K reported
in Table I were obtained by interpolating the results presented in
Fig. 1 of \cite{Shlomo2}. The corresponding values of the
temperature are also listed under the column T$_{F}$ and discussed
later in relation to Figs. 9b and 9c.
These temperature values  are in reasonable overall agreement with
measured temperatures of similar systems in this excitation energy
range, as systematically analyzed and presented in \cite{JBN1}.

In addition to the  above procedure to obtain the temperature of
fragmenting quasiprojectiles, we wish to investigate the possible
effect of expansion of the hot nucleus in the temperature
determination and, subsequently, in the extraction of the symmetry
energy coefficient.
For this purpose we employed the recently developed expanding
mononucleus model of Sobotka and T$\tilde{o}$ke \cite{Sobotka}.
Compared with the above non-expanding Fermi gas framework
\cite{Shlomo1}, this model incorporates the  expansion as a degree
of freedom \cite{Toke}. The effect of the variation the nucleon
effective mass in the level density is included  in a manner
analogous to \cite{Shlomo1}. For a given excitation energy
E$^{*}$/A, the collective expansion energy is taken into account.
The compressibility is chosen to correspond to a soft equation of
state.
The density of the mononucleus is determined so
that it maximizes the entropy. The temperature, in turn, is obtained
from the maximum entropy state. 
Using this model, we calculated the values of the temperature for
the quasiprojectiles of the reactions studied in this work which are
summarized in Table I under the column T$_{m}$ and later presented
in Fig. 9.
As we observe in Table I (and in Fig. 9b), the temperatures obtained
with this model are systematically lower that those obtained from
the (non-expanding) Fermi gas model. Despite differences in the
details of the two models, this difference may be understood
qualitatively as the effect of expansion: in the expanding
mononucleous  model, part of the available excitation energy is
allocated as potential energy of expansion, leaving the rest of the
amount as thermal energy and, thus, leading to lower temperature.
For completeness in the following discussion, both approaches for
the determination of the temperature will be employed, and the
respective results of the symmetry energy coefficient will be
presented and discussed.

\subsection{ Determination of the symmetry energy coefficient}
Having studied  the isoscaling behavior and determined the
corresponding excitation energies and temperatures, we currently
turn our discussion to the possibility of obtaining information on
the symmetry energy and its dependence on excitation energy. The key
quantity for this effort is the isoscaling parameter $\alpha$.
It has been shown \cite{Tsang1,Botvina,Ono03} that the isoscaling
parameter $\alpha$ is directly related to the coefficient C$_{sym}$
of the symmetry energy term $E_{sym} = C_{sym} (N-Z)^{2}/A $ of the
nuclear binding energy. The following relation has been obtained in
the framework of the grand canonical limit of the statistical
multifragmentation model (SMM) \cite{Botvina}, in the
expanding--emitting source (EES) model \cite{Tsang1,Tsang2} and in
the framework of dynamical calculations with the AMD model
\cite{Ono03,Ono04}:
\begin{equation}
     \alpha = 4 \, \frac {C_{sym}}{ T } ((\frac{Z}{A})^2_{1}  -  (\frac{Z}{A})^2_{2} )
\label{a1}
\end{equation}
where the atomic number Z and the mass number A  refer to the
fragmenting quasiprojectiles from reactions 1 and 2.

In principle,  Eq. \ref{a1} can serve as the basis for determining
the symmetry energy coefficient C$_{sym}$ for expanded
multifragmenting quasiprojectiles.
For this purpose, the isoscaling
parameter $\alpha$, the isotopic composition and the temperature of
fragmenting quasiprojectiles should be determined.
From the present study of heavy fragment isoscaling, the values of
the isoscaling parameter $\alpha$ have been extracted for each
reaction pair, as already discussed. The difference in isospin
asymmetry between quasiprojectiles of each reaction pair is taken to
be equal to that of the combined (fully mixed) system. We assumed
that N/Z equilibration has been reached in the present reactions at
the energy range 20--25 MeV/nucleon, as suggested  by recent
experimental studies \cite{GSnz} and calculations \cite{BAOnz}.  In
addition, in peripheral collisions at these bombarding energies, the
effect of pre-equilibrium emission does not appreciably change the
difference in isospin asymmetry, as has been concluded by
calculations using the model framework of \cite{MVnpa}. The same
conclusion has also been reached in the recent isoscaling studies of
spectator fragmentation at higher energies \cite{Trautmann}.
Finally, it should be noted that the effect of secondary decay on
the values of the isoscaling parameter
 $\alpha$ in the excitation energy range E$^{*}$/A $<$ 3 MeV is
 expected to be small \cite{Shetty05a}.
Thus, by using the experimental isoscaling parameter $\alpha$, the
temperature and the difference in isospin asymmetry obtained as
described above, we can obtain the values of the symmetry energy
coefficient C$_{sym}$ for each case corresponding  to the various
values of the excitation energy, as summarized in Table I. In the
following, we will discuss in further detail the steps involved to
extract C$_{sym}$ with the help of Figs. 8 and 9.


    \begin{figure}[h]                                        

    \includegraphics[width=0.47\textwidth, height=0.28\textheight ]{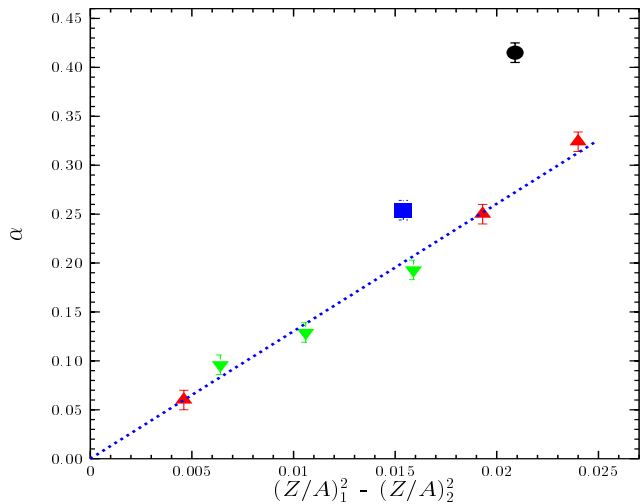}

    \caption{ (Color online)
           Isoscaling parameter $\alpha$ as a function of
           $ (Z/A)_{1}^{2} -  (Z/A)_{2}^{2} $ for each reaction pair
           studied (see text).
           Circle:  $^{86}$Kr + $^{124,112}$Sn.
           Square:  $^{86}$Kr + $^{64,58}$Ni.
           Upright triangles: $^{64}$Ni + $^{124,112}$Sn,
           $^{64,58}$Ni, $^{232}$Th--$^{208}$Pb, respectively in
           decreasing values of $\alpha$.
           Inverted triangles: $^{136}$Xe + $^{124,112}$Sn,
           $^{64,58}$Ni, $^{232}$Th--$^{197}$Au, respectively in
           decreasing values of $\alpha$.
           The straight dotted line indicates the linear
           relationship of the  $^{64}$Ni points (see text).
           }
    \label{a-systematics}
    \end{figure}

Fig. 8 presents the experimental values of the isoscaling parameter
$\alpha$ for the various systems studied in  this work as a function
of the difference in isospin composition  of each pair of reactions
expressed as $(\frac{Z}{A})^2_{1} - (\frac{Z}{A})^2_{2}$.
The various symbols correspond to the reactions studied as explained
in the caption of the figure. It is interesting to point out the
linear relationship  of the three points of the $^{64}$Ni data, all
of which have approximately the same excitation energy  of 2.9
MeV/nucleon. A similar observation can be made for the $^{136}$Xe
data points, which correspond to a common excitation energy of 2.5
MeV/nucleon.

Using the values of $\alpha$ and
$(\frac{Z}{A})^2_{1} - (\frac{Z}{A})^2_{2}$,
we obtain the parameter:
\begin{equation}
 c \equiv
  \frac{\alpha}{ (\frac{Z}{A})^2_{1}-(\frac{Z}{A})^2_{2} }
\label{c1}
\end{equation}
 which we call ``reduced'' isoscaling parameter. The values
of c with respect to the corresponding excitation energies are shown
in Fig. 9a. Because of the common excitation energy for the three
$^{64}$Ni  reaction pairs (and similarly for the three $^{136}$Xe
pairs), we plotted the corresponding average values of c in Fig. 9a.

Fig. 9b shows the temperatures calculated both with the Fermi gas
model (closed symbols) and the expanding mononucleus model (open
symbols). As already mentioned, the temperature values calculated
with the mononucleus expansion model are systematically lower (and
appear to lead to a plateau with respect to excitation energy, as
also discussed in \cite{Sobotka}).

 From Eq. \ref{a1}, with the definition of Eq. \ref{c1},
for the symmetry energy coefficient we simply have:
\begin{equation}
 C_{sym} = \frac{c\,T}{4}
\label{c2}
\end{equation}
The values of C$_{sym}$ obtained using this equation are shown in
Fig. 9c with closed  symbols (using the Fermi gas temperature) and
open symbols (using the expanding mononucleus temperature). As
previously discussed, the simple (non-expanding) Fermi gas
temperatures are in overall agreement with the existing experimental
systematics, thus the extracted values of C$_{sym}$ should be
considered the most appropriate ones for the present analysis.
Obviously, the values of the symmetry energy coefficient obtained
using the expanding mononucleus temperature are, as expected,
systematically lower. However, within both sets of C$_{sym}$ values,
a decreasing trend with increasing excitation energy is clearly
observed. In the following, only the values of C$_{sym}$
corresponding to the Fermi gas temperatures will be discussed
further.


    \begin{figure}[t]                                        

    \includegraphics[width=0.47\textwidth, height=0.53\textheight ]{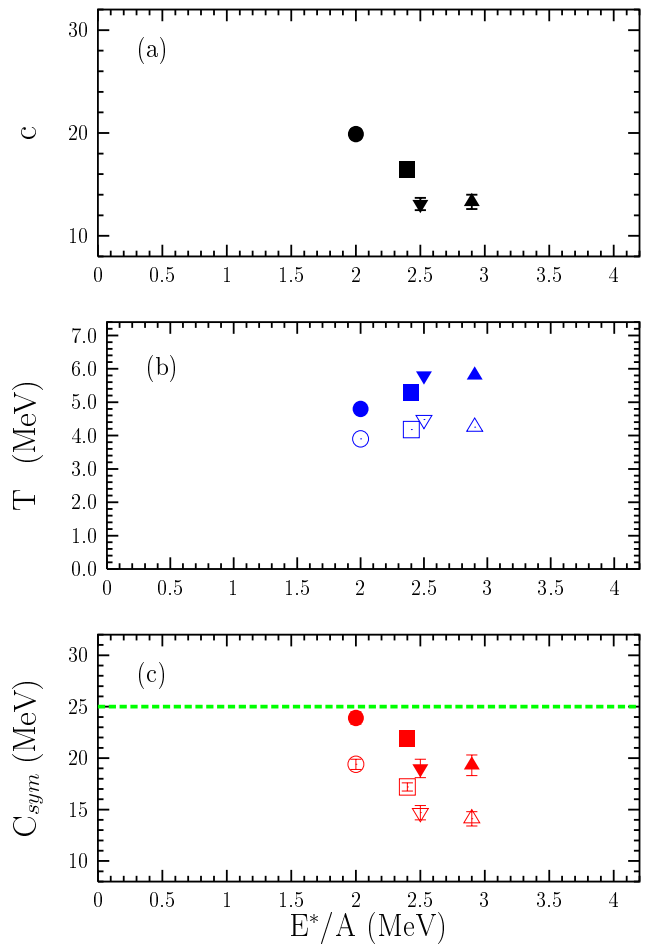}

    \caption{ (Color online) (a) Reduced isoscaling parameter
              $ c \equiv   \alpha / ( (Z/A)_{1}^{2} -  (Z/A)_{2}^{2} )$
              as a function 
              E$^{*}$/A.
              Circle:  $^{86}$Kr + $^{124,112}$Sn.
              Square:  $^{86}$Kr + $^{64,58}$Ni.
              Upright triangle:  $^{64}$Ni reactions.
              Inverted triangle: $^{136}$Xe  reactions.
              (b) Temperature as a function of E$^{*}$/A.
              Open symbols: Fermi gas. Closed symbols:
              expanding mononucleus  model \cite{Sobotka}. Symbol types as in (a).
              (c) Symmetry energy coefficient C$_{sym}$
              as a function of E$^{*}$/A.
              Open and closed symbols correspond to the two
              temperature estimates as in (b). Symbol types as in
              (a).
              The horizontal dashed  line indicates
              the typical value of C$_{sym}$ for cold nuclei.
             }
    \label{c}
    \end{figure}

From Fig. 9c (closed symbols), we observe that, at excitation energy
E$^{*}$/A $\sim$ 2.0 MeV, the value of the symmetry energy
coefficient is near (slightly lower than) the conventional value
C$_{sym,0} \simeq $ 25 MeV for cold (unexpanded) nuclei.
With increasing excitation energy, however, C$_{sym}$ appears to
decrease monotonically.
The observed  considerable decrease of the symmetry energy
coefficient with excitation energy towards the multifragmentation
regime is in overall qualitative agreement with the conclusions of
Shetty et al. \cite{Shetty05a}, Le F$\grave{e}$vre et al.
\cite{Trautmann} and Henzlova et al. \cite{ZKHS}.
We wish to point out, however, that the present study of heavy
residue isoscaling reveals the gradual decrease of the symmetry
energy C$_{sym}$ with increasing excitation energy in the range
E$^{*}$/A=2.0--2.9 MeV.


The observed decrease in the fragment symmetry energy with
increasing excitation has important consequences for the formation
of hot primary fragments and, as recently shown in \cite{Botvina1},
their subsequent decay.  Similar hot nuclei are also produced in the
interior of a collapsing star and subsequent supernova explosion
\cite{Botvina0,Botvina2}. In these works it is  predicted that a
small decrease in the symmetry energy coefficient can significantly
alter the elemental abundance and the synthesis of heavy elements in
type II supernova. In light of this intimate connection, the present
measurements and results can provide an important testing ground to
study the role of the symmetry energy in the formation and decay of
hot fragments and, subsequently,  to implement this knowledge in the
simulation of the distribution of hot exotic nuclei in supernova and
other hot and dense stellar environments. Along these lines, a
thorough comparison of the isoscaling properties and the N/Z
characteristics of the residues of the present work with the
statistical multifragmentation model (SMM) \cite{SMM} is currently
underway.

\subsection{ Final remarks and future directions}
As concluding remarks from this study, we point out that the present
mass spectrometric data provided information on the isoscaling
properties of heavy projectile fragments and, in parallel,
information on the average excitation energy of the primary
fragments (via residue velocity measurements). These two main
experimental observables  were used in the analysis presented in
this work to extract  the symmetry energy coefficient at the
corresponding excitation energy. An important advantage associated
with the detailed study of heavy residues is that, in the respective
excitation energy range E$^{*}$/A $<$ 3 MeV, the effect of
de-excitation on the isoscaling parameter $\alpha$ is expected to be
small, as discussed earlier,  thus slightly affecting the procedure
to obtain the symmetry energy coefficient.
As a future experimental step, we wish to propose the combination of
a mass separator/spectrometer with a multidetector system capable of
providing full acceptance and characterization for all fragments of
the decaying projectile.
Such an apparatus could enable the determination of the excitation
energy on an event-by-event basis, along with temperature
measurements via standard double-isotope techniques \cite{JBN1}.
This way, a detailed mapping of the excitation energy could be
performed for various reaction pairs covering the whole range from
low excitation energies (where only heavy residues are present that
can be separated and identified with the spectrometer) to the
multifragmentation region. At each excitation energy bin, the
isoscaling parameter $\alpha$ and the temperature can be obtained.
In addition, these measurements can be supplemented by efforts to
determine the density (e.g., following approaches as discussed in
\cite{JBN2} and \cite{Viola}). Consequently, a  correlation of the
symmetry energy coefficient with excitation energy and, possibly,
density may be obtained.
Finally, such advanced experimental studies can be extended to the
limits of isospin by taking advantage of present and future
developments of rare isotope beams.

\section{Summary and Conclusions}
The isoscaling properties of isotopically resolved projectile
residues from peripheral collisions  of $^{86}$Kr (25MeV/nucleon),
$^{64}$Ni (25MeV/nucleon) and $^{136}$Xe (20MeV/nucleon) beams  on
various target pairs are employed to probe the symmetry energy term
of the nuclear binding energy of hot fragments. The reactions of
$^{86}$Kr with $^{64,58}$Ni and $^{124,112}$Sn were studied with the
MARS recoil separator.  The reactions of  $^{64}$Ni  and  $^{136}$Xe
with $^{64,58}$Ni and $^{124,112}$Sn, as well as heavier targets
($^{197}$Au,  $^{208}$Pb, $^{232}$Th)  were studied  with the
Superconducting Solenoid Line (BigSol) at the Cyclotron Institute of
Texas A\&M University. The  present study focused on heavy
projectile fragments produced in peripheral and semiperipheral
collisions near the onset of multifragment emission (E$^{*}$/A =
2--3 MeV). For these fragments, the measured average velocities were
used to extract excitation energies. The excitation energies, in
turn, are employed to estimate the temperatures of the fragmenting
quasiprojectiles within the framework of the Fermi gas model. The
isoscaling analysis of the fragment yields provided the isoscaling
parameters $\alpha$ which, combined  with temperatures and isospin
asymmetries, provided the values of the symmetry energy. The
extracted value of the symmetry energy coefficient C$_{sym}$ at
E$^{*}$/A $\simeq$ 2 MeV is near (just below) the typical value
$\sim $25 MeV and is found to decrease monotonically  with further
increase of the excitation energy.  The observed decrease of
C$_{sym}$ with excitation energy is of significant importance to the
understanding of the formation and decay of hot nuclei not only in
nuclear multifragmentation, but also in supernova and other hot
stellar environments.

\section{Ackowledgements}
We wish to thank A. S. Botvina for fruitful and enlightening
discussions at various stages of this work. We are thankful to L. G.
Sobotka for providing us with his code on expanding mononucleus
calculations. We are grateful to F. D. Becchetti and T. W. O'Donnel
for valuable advise and discussions during the setup and
commissioning of the Superconducting Solenoid Line at the Cyclotron
Institute of Texas A\&M University. Finally, we wish to thank the
Cyclotron Institute staff for the excellent beam quality. This work
was supported in part by the Robert A. Welch Foundation through
grant No. A-1266, and the Department of Energy through grant No.
DE-FG03-93ER40773.  M.V. was also supported through grant
VEGA-2/5098/25 (Slovak Scientific Grant Agency).



\bibliography{csym}

\begin{thebibliography}{69}
\expandafter\ifx\csname natexlab\endcsname\relax\def\natexlab#1{#1}\fi
\expandafter\ifx\csname bibnamefont\endcsname\relax
  \def\bibnamefont#1{#1}\fi
\expandafter\ifx\csname bibfnamefont\endcsname\relax
  \def\bibfnamefont#1{#1}\fi
\expandafter\ifx\csname citenamefont\endcsname\relax
  \def\citenamefont#1{#1}\fi
\expandafter\ifx\csname url\endcsname\relax
  \def\url#1{\texttt{#1}}\fi
\expandafter\ifx\csname urlprefix\endcsname\relax\def\urlprefix{URL }\fi
\providecommand{\bibinfo}[2]{#2}
\providecommand{\eprint}[2][]{\url{#2}}

\bibitem[{\citenamefont{Steiner et~al.}(2005)\citenamefont{Steiner, Prakash,
  Lattimer, and Ellis}}]{ASTE05}
\bibinfo{author}{\bibfnamefont{A.~W.} \bibnamefont{Steiner}},
  \bibinfo{author}{\bibfnamefont{M.}~\bibnamefont{Prakash}},
  \bibinfo{author}{\bibfnamefont{J.~M.} \bibnamefont{Lattimer}},
  \bibnamefont{and} \bibinfo{author}{\bibfnamefont{P.}~\bibnamefont{Ellis}},
  \bibinfo{journal}{Phys. Rep.} \textbf{\bibinfo{volume}{411}},
  \bibinfo{pages}{325} (\bibinfo{year}{2005}).

\bibitem[{\citenamefont{Lee}(1996)}]{LEE96}
\bibinfo{author}{\bibfnamefont{C.}~\bibnamefont{Lee}}, \bibinfo{journal}{Phys.
  Rep.} \textbf{\bibinfo{volume}{275}}, \bibinfo{pages}{255}
  (\bibinfo{year}{1996}).

\bibitem[{\citenamefont{Bethe}(1990)}]{BET90}
\bibinfo{author}{\bibfnamefont{H.~A.} \bibnamefont{Bethe}},
  \bibinfo{journal}{Rev. Mod. Phys.} \textbf{\bibinfo{volume}{62}},
  \bibinfo{pages}{801} (\bibinfo{year}{1990}).

\bibitem[{\citenamefont{Pethick and Ravenhall}(1995)}]{PET95}
\bibinfo{author}{\bibfnamefont{C.~J.} \bibnamefont{Pethick}} \bibnamefont{and}
  \bibinfo{author}{\bibfnamefont{D.~G.} \bibnamefont{Ravenhall}},
  \bibinfo{journal}{Annu. Rev. Nucl. Part. Sci.} \textbf{\bibinfo{volume}{45}},
  \bibinfo{pages}{429} (\bibinfo{year}{1995}).

\bibitem[{\citenamefont{Lattimer and Prakash}(2000)}]{LAT00}
\bibinfo{author}{\bibfnamefont{J.~M.} \bibnamefont{Lattimer}} \bibnamefont{and}
  \bibinfo{author}{\bibfnamefont{M.}~\bibnamefont{Prakash}},
  \bibinfo{journal}{Phys. Rep.} \textbf{\bibinfo{volume}{333}},
  \bibinfo{pages}{121} (\bibinfo{year}{2000}).

\bibitem[{\citenamefont{Hix et~al.}(2003)}]{HIX03}
\bibinfo{author}{\bibfnamefont{W.~R.} \bibnamefont{Hix}} \bibnamefont{et~al.},
  \bibinfo{journal}{Phys. Rev. Lett.} \textbf{\bibinfo{volume}{91}},
  \bibinfo{pages}{201102} (\bibinfo{year}{2003}).

\bibitem[{\citenamefont{Stone et~al.}(2003)}]{JStone1}
\bibinfo{author}{\bibfnamefont{J.~R.} \bibnamefont{Stone}}
  \bibnamefont{et~al.}, \bibinfo{journal}{Phys. Rev. C}
  \textbf{\bibinfo{volume}{68}}, \bibinfo{pages}{034324}
  (\bibinfo{year}{2003}).

\bibitem[{\citenamefont{Botvina and Mishustin}(2004)}]{ZBotvina0}
\bibinfo{author}{\bibfnamefont{A.~S.} \bibnamefont{Botvina}} \bibnamefont{and}
  \bibinfo{author}{\bibfnamefont{I.~N.} \bibnamefont{Mishustin}},
  \bibinfo{journal}{Phys. Lett. B} \textbf{\bibinfo{volume}{584}},
  \bibinfo{pages}{233} (\bibinfo{year}{2004}).

\bibitem[{\citenamefont{Oyamatsu et~al.}(1998)}]{OYA98}
\bibinfo{author}{\bibfnamefont{K.}~\bibnamefont{Oyamatsu}}
  \bibnamefont{et~al.}, \bibinfo{journal}{Nucl. Phys. A}
  \textbf{\bibinfo{volume}{634}}, \bibinfo{pages}{3} (\bibinfo{year}{1998}).

\bibitem[{\citenamefont{Brown}(2000)}]{BRO00}
\bibinfo{author}{\bibfnamefont{B.~A.} \bibnamefont{Brown}},
  \bibinfo{journal}{Phys. Rev. Lett.} \textbf{\bibinfo{volume}{85}},
  \bibinfo{pages}{5296} (\bibinfo{year}{2000}).

\bibitem[{\citenamefont{Horowitz and Piekarewicz}(2001)}]{HOR01}
\bibinfo{author}{\bibfnamefont{C.~J.} \bibnamefont{Horowitz}} \bibnamefont{and}
  \bibinfo{author}{\bibfnamefont{J.}~\bibnamefont{Piekarewicz}},
  \bibinfo{journal}{Phys. Rev. Lett.} \textbf{\bibinfo{volume}{86}},
  \bibinfo{pages}{5647} (\bibinfo{year}{2001}).

\bibitem[{\citenamefont{Furnstahl}(2002)}]{FUR02}
\bibinfo{author}{\bibfnamefont{R.~J.} \bibnamefont{Furnstahl}},
  \bibinfo{journal}{Nucl. Phys. A} \textbf{\bibinfo{volume}{706}},
  \bibinfo{pages}{85} (\bibinfo{year}{2002}).

\bibitem[{\citenamefont{Khoa et~al.}(1996)\citenamefont{Khoa, von Oertzen, and
  Oglobin}}]{KHO96}
\bibinfo{author}{\bibfnamefont{D.~T.} \bibnamefont{Khoa}},
  \bibinfo{author}{\bibfnamefont{W.}~\bibnamefont{von Oertzen}},
  \bibnamefont{and} \bibinfo{author}{\bibfnamefont{A.~A.}
  \bibnamefont{Oglobin}}, \bibinfo{journal}{Nucl. Phys. A}
  \textbf{\bibinfo{volume}{602}}, \bibinfo{pages}{98} (\bibinfo{year}{1996}).

\bibitem[{\citenamefont{Zuo et~al.}(1999)\citenamefont{Zuo, Bombaci, and
  Lombardo}}]{ZUO99}
\bibinfo{author}{\bibfnamefont{W.}~\bibnamefont{Zuo}},
  \bibinfo{author}{\bibfnamefont{I.}~\bibnamefont{Bombaci}}, \bibnamefont{and}
  \bibinfo{author}{\bibfnamefont{U.}~\bibnamefont{Lombardo}},
  \bibinfo{journal}{Phys. Rev. C} \textbf{\bibinfo{volume}{60}},
  \bibinfo{pages}{024605} (\bibinfo{year}{1999}).

\bibitem[{\citenamefont{Brack et~al.}(1985)\citenamefont{Brack, Guet, and
  Hakansson}}]{BRA85}
\bibinfo{author}{\bibfnamefont{M.}~\bibnamefont{Brack}},
  \bibinfo{author}{\bibfnamefont{C.}~\bibnamefont{Guet}}, \bibnamefont{and}
  \bibinfo{author}{\bibfnamefont{H.~B.} \bibnamefont{Hakansson}},
  \bibinfo{journal}{Phys. Rep.} \textbf{\bibinfo{volume}{123}},
  \bibinfo{pages}{276} (\bibinfo{year}{1985}).

\bibitem[{\citenamefont{Pearson and Nayak}(2000)}]{PEA00}
\bibinfo{author}{\bibfnamefont{J.~M.} \bibnamefont{Pearson}} \bibnamefont{and}
  \bibinfo{author}{\bibfnamefont{R.~C.} \bibnamefont{Nayak}},
  \bibinfo{journal}{Nucl. Phys. A} \textbf{\bibinfo{volume}{668}},
  \bibinfo{pages}{163} (\bibinfo{year}{2000}).

\bibitem[{\citenamefont{Dieperink et~al.}(2003)}]{DIE03}
\bibinfo{author}{\bibfnamefont{A.~E.~L.} \bibnamefont{Dieperink}}
  \bibnamefont{et~al.}, \bibinfo{journal}{Phys. Rev. C}
  \textbf{\bibinfo{volume}{68}}, \bibinfo{pages}{064307}
  (\bibinfo{year}{2003}).

\bibitem[{\citenamefont{Myers and Swiatecki}(1966)}]{MYE66}
\bibinfo{author}{\bibfnamefont{W.~D.} \bibnamefont{Myers}} \bibnamefont{and}
  \bibinfo{author}{\bibfnamefont{W.~J.} \bibnamefont{Swiatecki}},
  \bibinfo{journal}{Nucl. Phys.} \textbf{\bibinfo{volume}{81}},
  \bibinfo{pages}{1} (\bibinfo{year}{1966}).

\bibitem[{\citenamefont{Pomorski and Dudek}(2003)}]{POM03}
\bibinfo{author}{\bibfnamefont{K.}~\bibnamefont{Pomorski}} \bibnamefont{and}
  \bibinfo{author}{\bibfnamefont{J.}~\bibnamefont{Dudek}},
  \bibinfo{journal}{Phys. Rev. C} \textbf{\bibinfo{volume}{67}},
  \bibinfo{pages}{044316} (\bibinfo{year}{2003}).

\bibitem[{\citenamefont{Stone et~al.}(2005)}]{JStone2}
\bibinfo{author}{\bibfnamefont{J.~R.} \bibnamefont{Stone}}
  \bibnamefont{et~al.}, \bibinfo{journal}{J. Phys. G, Nucl. Part. Phys.}
  \textbf{\bibinfo{volume}{31}}, \bibinfo{pages}{211} (\bibinfo{year}{2005}).

\bibitem[{\citenamefont{Lunney et~al.}(2003)\citenamefont{Lunney, Pearson, and
  Thibault}}]{Zmass-review}
\bibinfo{author}{\bibfnamefont{D.}~\bibnamefont{Lunney}},
  \bibinfo{author}{\bibfnamefont{J.~M.} \bibnamefont{Pearson}},
  \bibnamefont{and} \bibinfo{author}{\bibfnamefont{C.}~\bibnamefont{Thibault}},
  \bibinfo{journal}{Rev. Mod. Phys.} \textbf{\bibinfo{volume}{75}},
  \bibinfo{pages}{1021} (\bibinfo{year}{2003}).

\bibitem[{\citenamefont{Shetty et~al.}(2005{\natexlab{a}})\citenamefont{Shetty,
  Yennello, and Souliotis}}]{Shetty05}
\bibinfo{author}{\bibfnamefont{D.~V.} \bibnamefont{Shetty}},
  \bibinfo{author}{\bibfnamefont{S.~J.} \bibnamefont{Yennello}},
  \bibnamefont{and} \bibinfo{author}{\bibfnamefont{G.~A.}
  \bibnamefont{Souliotis}}, \bibinfo{journal}{nucl-ex/0505011}
  (\bibinfo{year}{2005}{\natexlab{a}}).

\bibitem[{\citenamefont{Shetty et~al.}(2004)}]{Shetty04}
\bibinfo{author}{\bibfnamefont{D.~V.} \bibnamefont{Shetty}}
  \bibnamefont{et~al.}, \bibinfo{journal}{Phys. Rev. C}
  \textbf{\bibinfo{volume}{70}}, \bibinfo{pages}{011601}
  (\bibinfo{year}{2004}).

\bibitem[{\citenamefont{Chen et~al.}(2005)\citenamefont{Chen, Ko, and
  Lee}}]{BAO05}
\bibinfo{author}{\bibfnamefont{L.~W.} \bibnamefont{Chen}},
  \bibinfo{author}{\bibfnamefont{C.~M.} \bibnamefont{Ko}}, \bibnamefont{and}
  \bibinfo{author}{\bibfnamefont{B.~A.} \bibnamefont{Lee}},
  \bibinfo{journal}{Phys. Rev. Lett.} \textbf{\bibinfo{volume}{94}},
  \bibinfo{pages}{032701} (\bibinfo{year}{2005}).

\bibitem[{\citenamefont{Khoa and Than}(2005)}]{KHO05}
\bibinfo{author}{\bibfnamefont{D.~T.} \bibnamefont{Khoa}} \bibnamefont{and}
  \bibinfo{author}{\bibfnamefont{H.~S.} \bibnamefont{Than}},
  \bibinfo{journal}{Phys. Rev. C.} \textbf{\bibinfo{volume}{71}},
  \bibinfo{pages}{044601} (\bibinfo{year}{2005}).

\bibitem[{\citenamefont{Li et~al.}(1998)\citenamefont{Li, Ko, and
  Bauer}}]{BAOreview}
\bibinfo{author}{\bibfnamefont{B.-A.} \bibnamefont{Li}},
  \bibinfo{author}{\bibfnamefont{C.~M.} \bibnamefont{Ko}}, \bibnamefont{and}
  \bibinfo{author}{\bibfnamefont{W.}~\bibnamefont{Bauer}},
  \bibinfo{journal}{Int. J. Mod. Phys. E} \textbf{\bibinfo{volume}{7}},
  \bibinfo{pages}{147} (\bibinfo{year}{1998}).

\bibitem[{\citenamefont{Li and Schroeder}(2001)}]{BAObook}
\bibinfo{editor}{\bibfnamefont{B.-A.} \bibnamefont{Li}} \bibnamefont{and}
  \bibinfo{editor}{\bibfnamefont{W.~U.} \bibnamefont{Schroeder}}, eds.,
  \emph{\bibinfo{title}{Isospin Physics in Heavy Ion Collisions at Intermediate
  Energies}} (\bibinfo{publisher}{Nova Science}, \bibinfo{address}{New York},
  \bibinfo{year}{2001}).

\bibitem[{\citenamefont{Ono et~al.}(2003)}]{Ono03}
\bibinfo{author}{\bibfnamefont{A.}~\bibnamefont{Ono}} \bibnamefont{et~al.},
  \bibinfo{journal}{Phys. Rev. C} \textbf{\bibinfo{volume}{68}},
  \bibinfo{pages}{051601} (\bibinfo{year}{2003}).

\bibitem[{\citenamefont{Zhang and Li}(2005)}]{ImQMD}
\bibinfo{author}{\bibfnamefont{Y.}~\bibnamefont{Zhang}} \bibnamefont{and}
  \bibinfo{author}{\bibfnamefont{Z.}~\bibnamefont{Li}}, \bibinfo{journal}{Phys.
  Rev. C} \textbf{\bibinfo{volume}{71}}, \bibinfo{pages}{024604}
  (\bibinfo{year}{2005}).

\bibitem[{\citenamefont{Tsang et~al.}(2001{\natexlab{a}})}]{Tsang1}
\bibinfo{author}{\bibfnamefont{M.~B.} \bibnamefont{Tsang}}
  \bibnamefont{et~al.}, \bibinfo{journal}{Phys. Rev. Lett.}
  \textbf{\bibinfo{volume}{86}}, \bibinfo{pages}{5023}
  (\bibinfo{year}{2001}{\natexlab{a}}).

\bibitem[{\citenamefont{Tsang et~al.}(2001{\natexlab{b}})}]{Tsang2}
\bibinfo{author}{\bibfnamefont{M.~B.} \bibnamefont{Tsang}}
  \bibnamefont{et~al.}, \bibinfo{journal}{Phys. Rev. C}
  \textbf{\bibinfo{volume}{64}}, \bibinfo{pages}{054615}
  (\bibinfo{year}{2001}{\natexlab{b}}).

\bibitem[{\citenamefont{Botvina et~al.}(2002)\citenamefont{Botvina, Lozhkin,
  and Trautmann}}]{Botvina}
\bibinfo{author}{\bibfnamefont{A.~S.} \bibnamefont{Botvina}},
  \bibinfo{author}{\bibfnamefont{O.~V.} \bibnamefont{Lozhkin}},
  \bibnamefont{and}
  \bibinfo{author}{\bibfnamefont{W.}~\bibnamefont{Trautmann}},
  \bibinfo{journal}{Phys. Rev. C} \textbf{\bibinfo{volume}{65}},
  \bibinfo{pages}{044610} (\bibinfo{year}{2002}).

\bibitem[{\citenamefont{Veselsky
  et~al.}(2004{\natexlab{a}})\citenamefont{Veselsky, Souliotis, and
  Jandel}}]{MVffiso}
\bibinfo{author}{\bibfnamefont{M.}~\bibnamefont{Veselsky}},
  \bibinfo{author}{\bibfnamefont{G.~A.} \bibnamefont{Souliotis}},
  \bibnamefont{and} \bibinfo{author}{\bibfnamefont{M.}~\bibnamefont{Jandel}},
  \bibinfo{journal}{Phys. Rev. C} \textbf{\bibinfo{volume}{69}},
  \bibinfo{pages}{044607} (\bibinfo{year}{2004}{\natexlab{a}}).

\bibitem[{\citenamefont{Friedman}(2004)}]{Friedman}
\bibinfo{author}{\bibfnamefont{W.~A.} \bibnamefont{Friedman}},
  \bibinfo{journal}{Phys. Rev. C} \textbf{\bibinfo{volume}{69}},
  \bibinfo{pages}{031601} (\bibinfo{year}{2004}).

\bibitem[{\citenamefont{Souliotis et~al.}(2003{\natexlab{a}})}]{GSiso}
\bibinfo{author}{\bibfnamefont{G.~A.} \bibnamefont{Souliotis}}
  \bibnamefont{et~al.}, \bibinfo{journal}{Phys. Rev. C}
  \textbf{\bibinfo{volume}{68}}, \bibinfo{pages}{024605}
  (\bibinfo{year}{2003}{\natexlab{a}}).

\bibitem[{\citenamefont{Souliotis et~al.}(2004)}]{GSnz}
\bibinfo{author}{\bibfnamefont{G.~A.} \bibnamefont{Souliotis}}
  \bibnamefont{et~al.}, \bibinfo{journal}{Phys. Lett. B}
  \textbf{\bibinfo{volume}{588}}, \bibinfo{pages}{35} (\bibinfo{year}{2004}).

\bibitem[{\citenamefont{Shetty et~al.}(2003)}]{Shetty03}
\bibinfo{author}{\bibfnamefont{D.~V.} \bibnamefont{Shetty}}
  \bibnamefont{et~al.}, \bibinfo{journal}{Phys. Rev. C}
  \textbf{\bibinfo{volume}{68}}, \bibinfo{pages}{021602}
  (\bibinfo{year}{2003}).

\bibitem[{\citenamefont{Veselsky
  et~al.}(2004{\natexlab{b}})\citenamefont{Veselsky, Souliotis, and
  Yennello}}]{MViso}
\bibinfo{author}{\bibfnamefont{M.}~\bibnamefont{Veselsky}},
  \bibinfo{author}{\bibfnamefont{G.~A.} \bibnamefont{Souliotis}},
  \bibnamefont{and} \bibinfo{author}{\bibfnamefont{S.~J.}
  \bibnamefont{Yennello}}, \bibinfo{journal}{Phys. Rev. C}
  \textbf{\bibinfo{volume}{69}}, \bibinfo{pages}{031602}
  (\bibinfo{year}{2004}{\natexlab{b}}).

\bibitem[{\citenamefont{Geraci et~al.}(2004)}]{Geraci}
\bibinfo{author}{\bibfnamefont{E.}~\bibnamefont{Geraci}} \bibnamefont{et~al.},
  \bibinfo{journal}{Nucl. Phys. A} \textbf{\bibinfo{volume}{732}},
  \bibinfo{pages}{173} (\bibinfo{year}{2004}).

\bibitem[{\citenamefont{LeF$\grave{e}$vre et~al.}(2005)}]{Trautmann}
\bibinfo{author}{\bibfnamefont{A.}~\bibnamefont{LeF$\grave{e}$vre}}
  \bibnamefont{et~al.}, \bibinfo{journal}{Phys. Rev. Lett.}
  \textbf{\bibinfo{volume}{94}}, \bibinfo{pages}{162701}
  (\bibinfo{year}{2005}).

\bibitem[{\citenamefont{Henzlova et~al.}(2005)}]{ZKHS}
\bibinfo{author}{\bibfnamefont{D.}~\bibnamefont{Henzlova}}
  \bibnamefont{et~al.}, \bibinfo{journal}{nucl-ex/0507003}
  (\bibinfo{year}{2005}).

\bibitem[{\citenamefont{Ono et~al.}(2004)}]{Ono04}
\bibinfo{author}{\bibfnamefont{A.}~\bibnamefont{Ono}} \bibnamefont{et~al.},
  \bibinfo{journal}{Phys. Rev. C} \textbf{\bibinfo{volume}{70}},
  \bibinfo{pages}{041604} (\bibinfo{year}{2004}).

\bibitem[{\citenamefont{Tribble et~al.}(1989)\citenamefont{Tribble, Burch, and
  Gagliardi}}]{MARS}
\bibinfo{author}{\bibfnamefont{R.~E.} \bibnamefont{Tribble}},
  \bibinfo{author}{\bibfnamefont{R.~H.} \bibnamefont{Burch}}, \bibnamefont{and}
  \bibinfo{author}{\bibfnamefont{C.~A.} \bibnamefont{Gagliardi}},
  \bibinfo{journal}{Nucl. Instrum. Methods Phys. Res. A}
  \textbf{\bibinfo{volume}{285}}, \bibinfo{pages}{441} (\bibinfo{year}{1989}).

\bibitem[{\citenamefont{Wilcke et~al.}(1980)}]{Wilcke}
\bibinfo{author}{\bibfnamefont{W.~W.} \bibnamefont{Wilcke}}
  \bibnamefont{et~al.}, \bibinfo{journal}{At. Data Nucl. Data Tables}
  \textbf{\bibinfo{volume}{25}}, \bibinfo{pages}{389} (\bibinfo{year}{1980}).

\bibitem[{\citenamefont{Souliotis et~al.}(2003{\natexlab{b}})}]{GSprl}
\bibinfo{author}{\bibfnamefont{G.~A.} \bibnamefont{Souliotis}}
  \bibnamefont{et~al.}, \bibinfo{journal}{Phys. Rev. Lett.}
  \textbf{\bibinfo{volume}{91}}, \bibinfo{pages}{022701}
  (\bibinfo{year}{2003}{\natexlab{b}}).

\bibitem[{\citenamefont{Souliotis et~al.}(2002)}]{GSplb}
\bibinfo{author}{\bibfnamefont{G.~A.} \bibnamefont{Souliotis}}
  \bibnamefont{et~al.}, \bibinfo{journal}{Phys. Lett. B}
  \textbf{\bibinfo{volume}{543}}, \bibinfo{pages}{163} (\bibinfo{year}{2002}).

\bibitem[{\citenamefont{Souliotis et~al.}(2003{\natexlab{c}})}]{GSnim}
\bibinfo{author}{\bibfnamefont{G.~A.} \bibnamefont{Souliotis}}
  \bibnamefont{et~al.}, \bibinfo{journal}{Nucl. Instrum. Methods Phys. Res. B}
  \textbf{\bibinfo{volume}{204}}, \bibinfo{pages}{166}
  (\bibinfo{year}{2003}{\natexlab{c}}).

\bibitem[{\citenamefont{O'Donnell}(2000)}]{Tom1}
\bibinfo{author}{\bibfnamefont{T.~W.} \bibnamefont{O'Donnell}},
  \bibinfo{journal}{Ph. D. Thesis, university of Michgan;
  http://www-personal.umich.edu/$\sim$twod/thesis/}  (\bibinfo{year}{2000}).

\bibitem[{\citenamefont{O'Donnell et~al.}(1999)}]{Tom2}
\bibinfo{author}{\bibfnamefont{T.~W.} \bibnamefont{O'Donnell}}
  \bibnamefont{et~al.}, \bibinfo{journal}{Nucl. Instrum. Methods Phys. Res. A}
  \textbf{\bibinfo{volume}{422}}, \bibinfo{pages}{513} (\bibinfo{year}{1999}).

\bibitem[{\citenamefont{Souliotis et~al.}(2005)}]{GS-bigsol}
\bibinfo{author}{\bibfnamefont{G.~A.} \bibnamefont{Souliotis}}
  \bibnamefont{et~al.}, \bibinfo{journal}{Progress in Research, Cyclotron
  Institute, Texas A\&M University (2003-2004), p. II-26; ibid (2002-2003) p.
  V-5; ibib (2001-2002) p. V-19; accessible at: http://cyclotron.tamu.edu}
  (\bibinfo{year}{2005}).

\bibitem[{\citenamefont{Geissel and Munzenberg}(1995)}]{spectrograph}
\bibinfo{author}{\bibfnamefont{H.}~\bibnamefont{Geissel}} \bibnamefont{and}
  \bibinfo{author}{\bibfnamefont{G.}~\bibnamefont{Munzenberg}},
  \bibinfo{journal}{Annu. Rev. Nucl. Part. Sci.} \textbf{\bibinfo{volume}{45}},
  \bibinfo{pages}{163} (\bibinfo{year}{1995}).

\bibitem[{\citenamefont{Moller et~al.}(1997)\citenamefont{Moller, Nix, and
  Kratz}}]{Moller}
\bibinfo{author}{\bibfnamefont{P.}~\bibnamefont{Moller}},
  \bibinfo{author}{\bibfnamefont{J.~R.} \bibnamefont{Nix}}, \bibnamefont{and}
  \bibinfo{author}{\bibfnamefont{K.~L.} \bibnamefont{Kratz}},
  \bibinfo{journal}{At. Data Nucl. Data Tables} \textbf{\bibinfo{volume}{66}},
  \bibinfo{pages}{131} (\bibinfo{year}{1997}).

\bibitem[{\citenamefont{Madani et~al.}(1996)}]{Madani}
\bibinfo{author}{\bibfnamefont{H.}~\bibnamefont{Madani}} \bibnamefont{et~al.},
  \bibinfo{journal}{Phys. Rev. C} \textbf{\bibinfo{volume}{54}},
  \bibinfo{pages}{1291} (\bibinfo{year}{1996}).

\bibitem[{\citenamefont{T$\tilde{o}$ke and Schroeder}(1992)}]{TokeEx}
\bibinfo{author}{\bibfnamefont{J.}~\bibnamefont{T$\tilde{o}$ke}}
  \bibnamefont{and} \bibinfo{author}{\bibfnamefont{W.~U.}
  \bibnamefont{Schroeder}}, \bibinfo{journal}{Annu. Rev. Nucl. Part. Sci.}
  \textbf{\bibinfo{volume}{42}}, \bibinfo{pages}{401} (\bibinfo{year}{1992}).

\bibitem[{\citenamefont{Veselsky et~al.}(2000)}]{MV_SiSn}
\bibinfo{author}{\bibfnamefont{M.}~\bibnamefont{Veselsky}}
  \bibnamefont{et~al.}, \bibinfo{journal}{Phys. Rev. C}
  \textbf{\bibinfo{volume}{62}}, \bibinfo{pages}{064613}
  (\bibinfo{year}{2000}).

\bibitem[{\citenamefont{Sobotka et~al.}(2004)}]{Sobotka}
\bibinfo{author}{\bibfnamefont{L.~G.} \bibnamefont{Sobotka}}
  \bibnamefont{et~al.}, \bibinfo{journal}{Phys. Rev. Lett.}
  \textbf{\bibinfo{volume}{93}}, \bibinfo{pages}{132702}
  (\bibinfo{year}{2004}).

\bibitem[{\citenamefont{Shlomo and Kolomietz}(2005)}]{Shlomo1}
\bibinfo{author}{\bibfnamefont{S.}~\bibnamefont{Shlomo}} \bibnamefont{and}
  \bibinfo{author}{\bibfnamefont{V.~M.} \bibnamefont{Kolomietz}},
  \bibinfo{journal}{Rep. Prog. Phys.} \textbf{\bibinfo{volume}{68}},
  \bibinfo{pages}{1} (\bibinfo{year}{2005}).

\bibitem[{\citenamefont{Natowitz et~al.}(2002{\natexlab{a}})}]{JBN1}
\bibinfo{author}{\bibfnamefont{J.~B.} \bibnamefont{Natowitz}}
  \bibnamefont{et~al.}, \bibinfo{journal}{Phys. Rev. C}
  \textbf{\bibinfo{volume}{65}}, \bibinfo{pages}{034618}
  (\bibinfo{year}{2002}{\natexlab{a}}).

\bibitem[{\citenamefont{Shlomo and Natowitz}(1991)}]{Shlomo2}
\bibinfo{author}{\bibfnamefont{S.}~\bibnamefont{Shlomo}} \bibnamefont{and}
  \bibinfo{author}{\bibfnamefont{J.~B.} \bibnamefont{Natowitz}},
  \bibinfo{journal}{Phys. Rev. C} \textbf{\bibinfo{volume}{44}},
  \bibinfo{pages}{2878} (\bibinfo{year}{1991}).

\bibitem[{\citenamefont{Shlomo and Natowitz}(1990)}]{Shlomo3}
\bibinfo{author}{\bibfnamefont{S.}~\bibnamefont{Shlomo}} \bibnamefont{and}
  \bibinfo{author}{\bibfnamefont{J.~B.} \bibnamefont{Natowitz}},
  \bibinfo{journal}{Phys. Lett. B} \textbf{\bibinfo{volume}{252}},
  \bibinfo{pages}{187} (\bibinfo{year}{1990}).

\bibitem[{\citenamefont{T$\tilde{o}$ke
  et~al.}(2003)\citenamefont{T$\tilde{o}$ke, Lu, and Schroeder}}]{Toke}
\bibinfo{author}{\bibfnamefont{J.}~\bibnamefont{T$\tilde{o}$ke}},
  \bibinfo{author}{\bibfnamefont{J.}~\bibnamefont{Lu}}, \bibnamefont{and}
  \bibinfo{author}{\bibfnamefont{W.~U.} \bibnamefont{Schroeder}},
  \bibinfo{journal}{Phys. Rev. C} \textbf{\bibinfo{volume}{67}},
  \bibinfo{pages}{034609} (\bibinfo{year}{2003}).

\bibitem[{\citenamefont{Li and Yennello}(1995)}]{BAOnz}
\bibinfo{author}{\bibfnamefont{B.-A.} \bibnamefont{Li}} \bibnamefont{and}
  \bibinfo{author}{\bibfnamefont{S.~J.} \bibnamefont{Yennello}},
  \bibinfo{journal}{Phys. Rev. C} \textbf{\bibinfo{volume}{52}},
  \bibinfo{pages}{1746} (\bibinfo{year}{1995}).

\bibitem[{\citenamefont{Veselsky}(2002)}]{MVnpa}
\bibinfo{author}{\bibfnamefont{M.}~\bibnamefont{Veselsky}},
  \bibinfo{journal}{Nucl. Phys. A} \textbf{\bibinfo{volume}{705}},
  \bibinfo{pages}{193} (\bibinfo{year}{2002}).

\bibitem[{\citenamefont{Shetty et~al.}(2005{\natexlab{b}})}]{Shetty05a}
\bibinfo{author}{\bibfnamefont{D.~V.} \bibnamefont{Shetty}}
  \bibnamefont{et~al.}, \bibinfo{journal}{Phys. Rev. C}
  \textbf{\bibinfo{volume}{71}}, \bibinfo{pages}{024602}
  (\bibinfo{year}{2005}{\natexlab{b}}).

\bibitem[{\citenamefont{Buyukcizmeci et~al.}(2005)\citenamefont{Buyukcizmeci,
  Ogul, and Botvina}}]{Botvina1}
\bibinfo{author}{\bibfnamefont{N.}~\bibnamefont{Buyukcizmeci}},
  \bibinfo{author}{\bibfnamefont{R.}~\bibnamefont{Ogul}}, \bibnamefont{and}
  \bibinfo{author}{\bibfnamefont{A.~S.} \bibnamefont{Botvina}},
  \bibinfo{journal}{Eur. Phys. Jou. A} \textbf{\bibinfo{volume}{25}},
  \bibinfo{pages}{57} (\bibinfo{year}{2005}).

\bibitem[{\citenamefont{Botvina and Mishustin}(2005)}]{Botvina2}
\bibinfo{author}{\bibfnamefont{A.~S.} \bibnamefont{Botvina}} \bibnamefont{and}
  \bibinfo{author}{\bibfnamefont{I.~N.} \bibnamefont{Mishustin}},
  \bibinfo{journal}{Phys. Rev. C} \textbf{\bibinfo{volume}{72}},
  \bibinfo{pages}{048801} (\bibinfo{year}{2005}).

\bibitem[{\citenamefont{Bondorf et~al.}(1995)}]{SMM}
\bibinfo{author}{\bibfnamefont{J.~P.} \bibnamefont{Bondorf}}
  \bibnamefont{et~al.}, \bibinfo{journal}{Phys. Rep.}
  \textbf{\bibinfo{volume}{257}}, \bibinfo{pages}{133} (\bibinfo{year}{1995}).

\bibitem[{\citenamefont{Natowitz et~al.}(2002{\natexlab{b}})}]{JBN2}
\bibinfo{author}{\bibfnamefont{J.~B.} \bibnamefont{Natowitz}}
  \bibnamefont{et~al.}, \bibinfo{journal}{Phys. Rev. C}
  \textbf{\bibinfo{volume}{66}}, \bibinfo{pages}{031601}
  (\bibinfo{year}{2002}{\natexlab{b}}).

\bibitem[{\citenamefont{Viola et~al.}(2004)}]{Viola}
\bibinfo{author}{\bibfnamefont{V.}~\bibnamefont{Viola}} \bibnamefont{et~al.},
  \bibinfo{journal}{Phys. Rev. Lett.} \textbf{\bibinfo{volume}{93}},
  \bibinfo{pages}{132701} (\bibinfo{year}{2004}).

\end{thebibliography}



\end{document}